\documentclass{elsart}

\begin{document}
\begin{frontmatter}
\title{BTeV Level 1 Vertex Trigger}
\author{Michael H.L.S. Wang}
\collab{For the BTeV Collaboration}
\address{Fermi National Accelerator Laboratory, Batavia, IL 60510, USA}

\begin{abstract} 
BTeV is a $B$-physics experiment that expects to begin collecting data at the C$0$ interaction region of the Fermilab Tevatron in the year 2006.  Its primary goal is  to achieve unprecedented levels of sensitivity in the study of CP violation, mixing, and rare decays in $b$ and $c$ quark systems.  In order to realize this, it will employ a state-of-the-art first-level vertex trigger (Level 1) that will look at every beam crossing to identify detached secondary vertices that provide evidence for heavy quark decays.  This talk will briefly describe the BTeV detector and trigger, focus on the software and hardware aspects of the Level 1 vertex trigger, and describe work currently being done in these areas.
\end{abstract}
\end{frontmatter}

\section{BTeV Detector}

The BTeV detector is a collider detector with two back-to-back forward spectrometers ($1.5<|\eta|<4.5$)  having sub-detectors in each arm for charged particle tracking, EM calorimetry, RICH particle ID and muon detection \cite{prop}.  At the core of the detector is a 30 station Si-pixel inner tracker spanning $\sim$120 cm centered at the IR ($\sigma_z$ = 30 cm) and immersed in a 1.6 Tesla dipole field.  There are over $20\times10^6$ active rectangular pixels measuring 50 $\times$ 400 $\mu \mathrm{m}^2$.  Each pixel station has two planes, one with narrow pixel dimension oriented in the $x$-direction and the other with narrow dimension in the $y$-direction.

\section{BTeV Trigger}

BTeV will operate at a luminosity of $2\times10^{32}$ cm$^{-2}$ s$^{-1}$ corresponding to $\langle2\rangle$ interactions per beam crossing at a crossing rate of 7.6 MHz.  Average event sizes will be $\sim$200 KB after zero-suppression of data is performed on-the-fly 
%%%%%%%%%%%%%%%%%%

\input epsf
\begin{figure}[ht]
\begin{center}
\leavevmode
\hbox{%
\epsfxsize=5.0in
\epsfbox{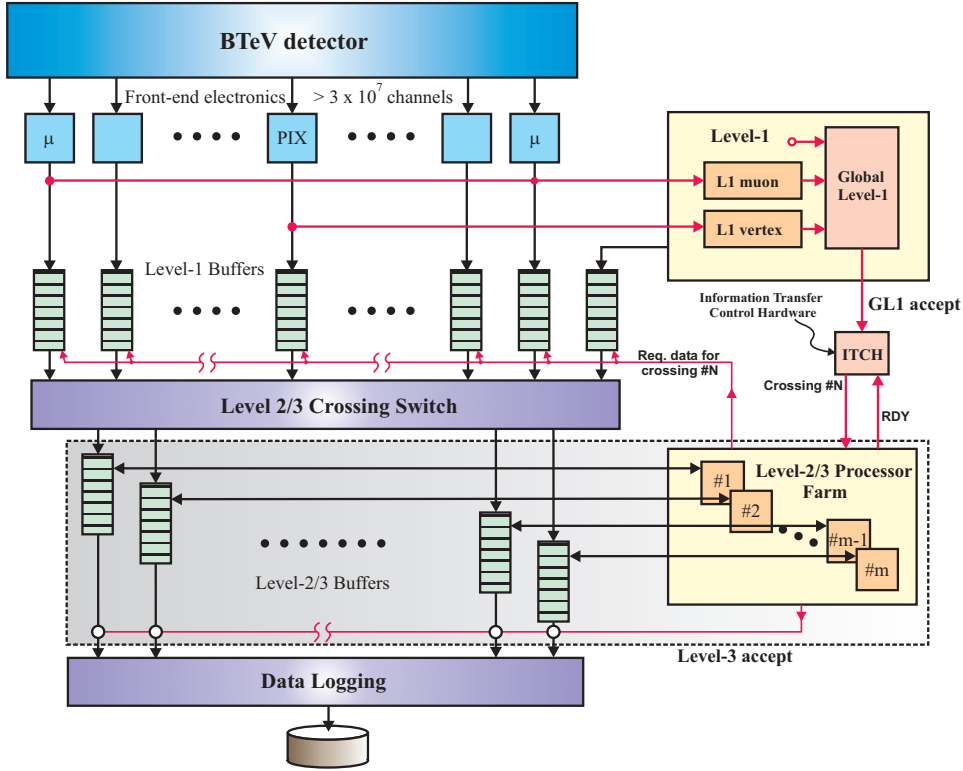}}
\caption{
\label{btevtrig}
\small BTeV Three-Level Trigger Architecture}
\end{center}
\end{figure}

%%%%%%%%%%%%%%%%%%
by front-end detector electronics.  Since every beam crossing will be processed, this translates to an extremely high data rate of $\sim$1.5 TB/sec!

To handle this high rate, BTeV will employ a three-level hierarchical trigger architecture \cite{paul} (see Fig. \ref{btevtrig}).  Sparsified data from all detector components will be sent via optical links to Level 1 (L1) buffers. Data from the pixel and muon detectors will also be sent to the  L1 trigger processor. Trigger decisions from the L1 muon and vertex trigger will be passed on to the Global Level 1 (GL1) trigger manager whose decisions will be stored as a list of accepted crossing numbers by the Information Transfer Control Hardware (ITCH).  L1 will reject ~99\% of all incoming events reducing the data rate by a factor of $\sim$100 to $\sim$15 GB/sec.

Levels 2 \& 3 (L2/3) will be implemented with a cluster of commodity CPU nodes.  A request from an idle L2 node will be sent to the ITCH which will respond by assigning an accepted crossing number to that node.  This node will then request a subset of the data (pixels+forward tracking) for that crossing from the L1 buffer managers.  A switch will combine pixel data with forward tracking data for the same crossing and route them to the buffer of an L2 node allowing a more refined analysis that will further reduce the data rate by a factor of $\sim$10.

If that passes the L2 selection criteria, the same processing node will then enter the L3 phase and request data from the rest of the sub-detectors be transferred from the L1 to the L3 buffers (L2/3 buffers will simply be the RAM attached to the processing node).  Using complete information from the detector, L3 will reduce the crossing rate by an additional factor of $\sim$2.  Since L3 will also be able to further compress the data in the accepted crossings by a factor of $\sim$4, the expected data rate out of L3 is $\sim$200 MB/sec.

\section{Level 1 Vertex Trigger Algorithm}

The first phase of the L1 vertex trigger algorithm is the track finding phase \cite{erik} which starts by finding the beginning and ending segments of tracks in two separate regions of the pixel planes, an inner region close to the beam axis and an outer region close to the edge of the pixel planes.  The search for the beginning and ending segments of tracks is restricted to these inner and outer regions respectively.  Segments are found using hit clusters from three adjacent pixel stations in the defined regions.  Inner segments are required to point back to the beam axis while outer segments are required to project outside pixel plane boundaries.  Once these segments are found, they are then matched to form complete tracks in the segment matching stage.

After complete tracks are found, the vertexing phase \cite{isik1,isik2} determines the momentum of each track and calculates its transverse distance from the beam axis.  Primary vertices are found by looping through all tracks with transverse momentum $p_T\le1.2$ GeV/c that appear to originate close to the beamline.
Remaining tracks are then tested for their detachment from the primary vertices found.  A L1 vertex trigger is generated if there are at least 2 tracks in the same arm of the BTeV detector satisfying the following criteria:  $p_T^2\ge0.25$ (GeV/c)$^2$, $b\ge6 \sigma$, and $b\le2$mm where $b$ is the impact parameter.

Our studies indicate that the L1 vertex trigger is able to reject 99\% of all minimum-bias events while accepting  $\sim$60-70\%  of the $B$-events that would survive our offline analysis cuts.

\section{Level 1 Vertex Trigger Hardware}

\begin{figure}[ht]
\begin{center}
\leavevmode
\hbox{%
\epsfxsize=5.0in
\epsfbox{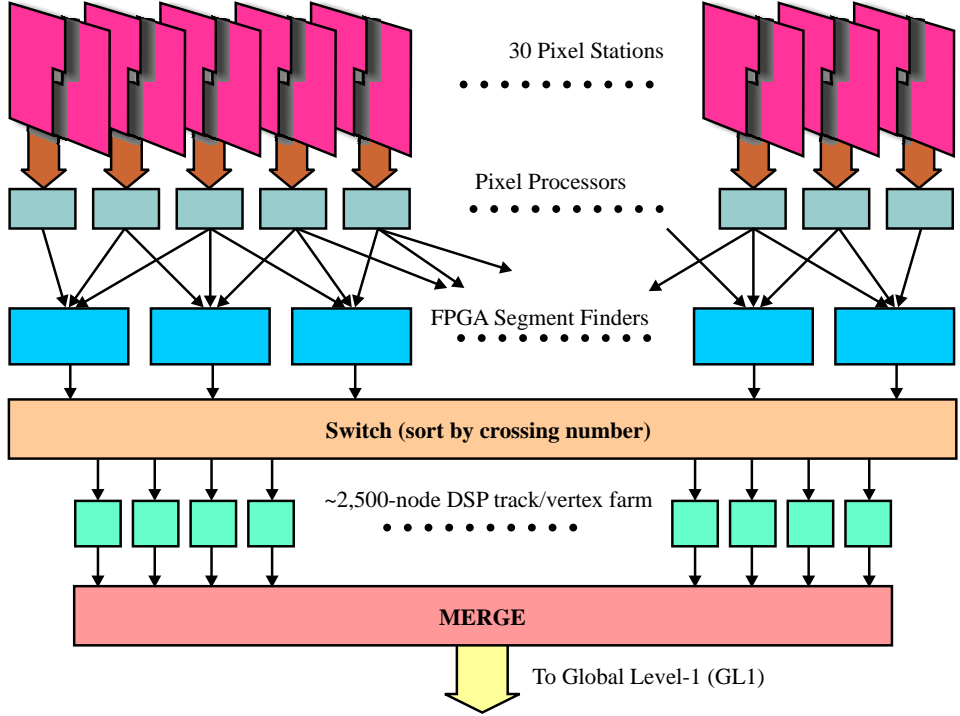}}
\caption{
\label{l1vtx}
\small BTeV Level-1 vertex trigger.}
\end{center}
\end{figure}

A block diagram of the L1 vertex trigger is shown in Fig. \ref{l1vtx}. Data from all 30 stations of the pixel detector are sent to FPGA based pixel processors that group individual pixel hits into clusters.  Hit clusters from three neighboring pixel stations are routed to FPGA based hardware that find beginning and ending segments of tracks in the track finding phase.
%%%%%%%%%%%%%%%%%%

\input epsf

%%%%%%%%%%%%%%%%%%
Segments found in this stage are sorted by a switch according to crossing
number and  routed to a DSP in the track/vertex farm.  This DSP performs the segment matching step of the track finding phase and the entire vertex finding phase.  Based on initial studies done for the BTeV proposal \cite{prop}, we estimate the average processing time for the combined segment matching plus vertex finding phase to take $\sim$330 $\mu$s on a single 167 MHz TI TMS320C6711 floating point DSP. Since the time between beam crossings is 132ns, this will require a total of $\sim$2,500 DSP in the track/vertex farm in order to examine every beam crossing.

\section{Work in Progress}

Since hand-optimized assembly code can be difficult to maintain and debug, we are working on an optimized C version of the L1 segment matching and vertexing code.  A
version is currently running on a C6711 DSK board in which we have replaced
costly function calls with intrinsics that map directly to built-in DSP
instructions and completely rewritten major sections of the code.  Preliminary results
indicate that this optimized C version is only a factor of $\sim$4-5 slower than
the hand-optimized assembly version used for initial timing estimates.  While we expect substantial gains in DSP chip 
performance in the next few years, we will continue to explore ways to improve code performance, resorting to hand-optimized assembly programming only when
necessary.

We are developing a prototype board for the segment matching and vertexing portion of the L1 trigger with four DSP's on mezzanine cards.  Onboard FPGA input and output buffers will have an LVDS interface to a host computer through which simulated data can be sent.  Buffer managers in the FPGA will channel data to and from each DSP via DMA transfers. Trigger results from each processor will be sent through an FPGA interface to an on-board $\mu$-controller that will format the results and forward them to GL1 through ArcNet. A host computer serving as a pixel trigger supervisor monitor will be connected to a second on-board $\mu$-controller through ArcNet, allowing commands to be sent to the board, initialization of the DSP's, and providing hardware monitoring and fault detection.  JTAG ports will also be available for real-time debugging and initial start-up of the prototype.

We are also implementing the L1 segment finding algorithm in FPGA's and doing simulations and queuing studies of the algorithm with MATLAB.

As a final note BTeV will greatly benefit from a newly approved NSF funded project called RTES (Real Time Embedded Systems) \cite{rtes} which will develop a semi-autonomous, self-monitoring, fault-tolerant and adaptive framework for addressing issues facing complex computing architectures like that of BTeV.

\section{Conclusion}

We have designed a Level 1 vertex trigger that will address the demanding requirements of BTeV.  We are currently working on optimizing a high-level version of the vertex trigger algorithm and developing a prototype board for investigating various aspects of the L1 vertex trigger hardware.


\begin{thebibliography}{7}

\bibitem{prop}  A. Kulyavtsev et al., BTeV proposal, Fermilab, May 2000.
\bibitem{paul}P. Lebrun,  The BTeV triggers and data acqusition systems, talk presented at the session on New Detector Technologies, at the meeting of the Division of Particles and Fields of The American Physical Society, Columbus, OH, August 9-12 2000.
\bibitem{erik} E.E. Gottschalk, BTeV detached vertex trigger, Nucl. Instrum. Meth. A 473 (2001) 167.
\bibitem{isik1} R. Isik, Real-time pattern-recognition for HEP, University of Pennsylvania, preprint UPR-233E, 1996.
\bibitem{isik2} R. Isik, W. Selove, K. Sterner, Monte Carlo results for a secondary-vertex trigger with on-line tracking, University of Pennsylvania, preprint UPR-234E, 1996.
\bibitem{rtes} E.E. Gottschalk, The BTeV DAQ and trigger system- some throughput, usability and fault tolerance aspects, Proc. CHEP 2001, Beijing, Sept. 3-7 2001, p. 628.

\end{thebibliography}
\end{document}